\documentclass[intlimits,twoside,a4paper]{article}

\usepackage{amsmath,amssymb}
\usepackage{graphicx}
\usepackage{wrapfig}

\usepackage[T2A]{fontenc}
\usepackage[cp1251]{inputenc}

\usepackage{epsf,multicol,ifthen}

\usepackage[eqsecnum]{cmpj3}

\issue{2018}{21}{2}{23701}
\doinumber{10.5488/CMP.21.23701}

\title[The theory of electron states on the dynamically deformed adsorbed surface of a solid]%
{The theory of electron states on the dynamically deformed adsorbed surface of a solid}
\author[R.M.~Peleshchak, M.Ya.~Seneta ]{R.M.~Peleshchak, M.Ya.~Seneta\footnote{E-mail: marsen18@i.ua}\, }
\address{Drohobych Ivan Franko State Pedagogical University, 24 Franko St., 82100 Drohobych, Ukraine}

\date{Received January 12, 2018, in final form March 23, 2018}

\begin{document}

\maketitle

\begin{abstract}
Dispersion relations for the spectra of surface electron states on a dynamically deformed adsorbed surface of a monocrystal with the Zinc blende structure are received. It is established that the dependences of the band gap width and of the concentration of electrons on the concentration of adatoms ${{N}_{0\text{d}}}$  upon the solid surface are of nonmonotonous character.

\pacs 73.20.At, 73.0.Hb, 68.60.Bs
\end{abstract}

\section{Introduction}
The creation of a new class of microelectronic and nanoelectronic devices with controlled pa\-ra\-me\-ters needs a research of the excitation mechanisms of electronic states upon the adsorbed surface of semiconductors. One of these mechanisms of excitation is a dynamic deformation in the subsurface layer of a solid. Such a dynamic deformation upon the adsorbed surface of a solid can be formed by a quasi-Rayleigh acoustic wave~\cite{Liu16}. Interaction between the quasi-Rayleigh acoustic wave and adatoms renormalizes the spectrum of the surface electron states due to the deformation potential~\cite{Bir74}. Tech\-no\-lo\-gically changing the concentration of adsorbed atoms, it is possible to change the frequency of a surface acoustic wave~(SAW) and the electronic structure of a subsurface layer. Such a correlation between the concentration of adsorbed atoms, the frequency of a surface acoustic wave and the electron structure of a subsurface layer can be used in practice for the change of coefficients of electromagnetic waves reflection from the interface of the media and for the change of a dispersion law of plasma oscillations~\cite{Kha02}.

At present, there are many works~\cite{Kha02,Pog93, Kha11, Kha07, Yak08} devoted to the investigation of surface quantum electronic states. At the same time, the main focus was concentrated on the study of electron states upon the crystal surface, due to the breakdown of periodic crystalline potential. It is known that if the surface is smooth, then the surface electron states do not arise.

In particular, the authors of the work~\cite{Kha02} investigated the surface electron states of a semiconductor bounded by an uneven surface having an infinitely high potential barrier. The surface of a semiconductor was considered without adsorbed atoms, and the surface roughness was formed by a quasi-Rayleigh acoustic wave.

The influence of interaction between adatoms and the self-consistent acoustic quasi-Rayleigh wave on its dispersion and on the width of the phonon mode at various values of concentration of adatoms was investigated in the works~\cite{Pel14,Pel17}.

In the work~\cite{Sen17} within the long-wave approximation, taking into account the nonlocal elastic interaction between the adsorbed atom and the matrix atoms and the mirror image forces, the deformation potential of the acoustic quasi-Rayleigh wave is investigated, and the dependences of the deformation potential amplitude and of the surface roughness height on the concentration of adsorbed atoms are calculated.

The conditions of the appearance of localized electron states upon the semiconductor surface with inequalities formed by the adsorbed atoms and by the acoustic quasi-Rayleigh wave were investigated in the work~\cite{Sene17}.

The purpose of this work is to investigate the influence of the concentration of adsorbed atoms on the spectrum of the surface electron states and on the distribution of electron density upon a dynamically deformed adsorbed surface of a monocrystal with the Zinc blende structure.

\section{Formulation of the problem}

Here, we are considering the subsurface layer of a cubic crystal in a molecular beam epitaxy or implantation process~\cite{Pel14}. Under the action of a flow of atoms, the atoms are adsorbed with an average concentration ${{N}_{0\text{d}}}$. Due to the deformation potential and the local renormalization of the surface energy, the adatoms and the deformation field of the surface acoustic quasi-Rayleigh wave inhomogeneously deform the subsurface layer. In its turn, by way of the deformation potential the self-consistent deformation redistributes the adsorbed atoms along the surface. The influence of adsorbed atoms is reduced to a change of boundary conditions for a stress tensor ${{\sigma }_{ij}}$ on $z=0$ surface. The layer, being defect-enriched by adatoms,  can be considered as a film of $a$ thickness, $\rho$ density and Young's modulus $E$. It is rigidly connected with a substrate, the other monocrystal part, having elastic parameters ${{\rho }_\text{s}}$ and ${{E}_\text{s}}$. The connecting plane of the film and substrate $z=a$ is parallel to the surface (100). The $z$-axis  is directed into the single crystal depth, the axes \textit{x} and \textit{y} are directed along two orthogonal crystallographic directions of type [100].
A surface acoustic quasi-Rayleigh wave, extending in $x$-axis direction with ${\omega }'(\vec{q},{{N}_{0\text{d}}})$ frequency, forms a dynamic deformation and interacts with the adatoms. The deformed surface form along the $x$-axis, depending on time, can be described by the following function~\cite{Sene17}:
\begin{equation}
\label{eq1}
{{z}_{0}}(x,t)=\tilde{\varsigma }({{N}_{0\text{d}}})\cos \left[qx-{\omega }'(q,{{N}_{0\text{d}}})\cdot t\right],		
\end{equation}
where $q=\frac{2\piup }{{{L}_{x}}}$, ${{L}_{x}}$ is the period of roughnesses (the length of acoustic wave) along the $x$-axis. The dispersion law ${\omega }'(q,{{N}_{0\text{d}}})=\operatorname{Re}\omega $ of a quasi-Rayleigh wave and the width ${\omega }''(q,{{N}_{0\text{d}}})=\operatorname{Im}\omega $ of the acoustic phonon mode are determined ~\cite{Pel16}:
\begin{eqnarray}
\label{eq2}
  &&{\omega }'(q,{{N}_{0\text{d}}})={{c}_\text{t}}q{{\xi }_{0}}\left( 1-\frac{1}{{f}'({{\xi }_{0}})}\frac{2{{\xi }_{0}}{{\theta }_\text{d}}{{N}_{0\text{d}}}}{{{k}_\text{B}}T\rho c_\text{l}^{2}}\cdot \frac{{{D}_\text{d}}^{2}q \left[1-\frac{2}{3}\frac{1-2\nu }{K(1-\nu )a}\frac{{{\theta }_\text{d}}^{2}}{{{k}_\text{B}}T}{{N}_{0\text{d}}}\right]q}{{{\Big\{ {{D}_\text{d}}\left[1-\frac{2}{3}\frac{1-2\nu }{K(1-\nu )a}\frac{{{\theta }_\text{d}}^{2}}{{{k}_\text{B}}T}{{N}_{0\text{d}}}\right] \Big\}}^{2}}{{q}^{2}}+c_\text{t}^{2}{{\xi }_{0}}^{2}} \right. \nonumber \\
  &&\left. \times \left(1-l_\text{d}^{2}{{q}^{2}}\right)\left[ q\sqrt{1-{{\xi }_{0}}^{2}}\frac{\partial F}{\partial {{N}_{1\text{d}}}}+\left(2-{{\xi }_{0}}^{2}\right)\frac{{{\theta }_\text{d}}}{2a} \right] \right), 
\end{eqnarray}
\begin{eqnarray}
\label{eq3}
  &&{\omega }''(q,{{N}_{0\text{d}}})=-{{c}_\text{t}}q\frac{1}{{f}'({{\xi }_{0}})}\frac{2{{\xi }^{2}}{{\theta }_\text{d}}{{N}_{0\text{d}}}}{{{k}_\text{B}}T\rho c_\text{l}^{2}}\cdot \frac{{{D}_\text{d}}q}{{{\Big\{ {{D}_\text{d}}\left[1-\frac{2}{3}\frac{1-2\nu }{K(1-\nu )a}\frac{{{\theta }_\text{d}}^{2}}{{{k}_\text{B}}T}{{N}_{0\text{d}}}\right] \Big\}}^{2}}{{q}^{2}}+c_\text{t}^{2}{{\xi }_{0}}^{2}} \nonumber \\
  &&\times \left(1-l_\text{d}^{2}{{q}^{2}}\right)\left[ q\sqrt{1-{{\xi }_{0}}^{2}}\frac{\partial F}{\partial {{N}_{\text{d}1}}}+\left(2-{{\xi }_{0}}^{2}\right)\frac{{{\theta }_\text{d}}}{2a} \right], 
\end{eqnarray}
where ${{N}_{0\text{d}}}$ is a the spatially homogeneous component of the concentratoin of adatoms~\cite{Pel17}; $K$ is the modulus of elasticity; $l_\text{d}^{2}$  is the average of the square of a characteristic distance of the interaction between the adatom and the matrix atoms; $\nu $ is the Poisson coefficient; ${{\theta }_\text{d}}$ is the surface deformation potential~\cite{Sen17}; ${{D}_\text{d}}$ is the diffusion coefficient of the adatom; \textit{T} is the temperature of a substrate; ${{k}_\text{B}}$ is the Boltzmann constant.

The height of the roughness $\tilde{\varsigma }({{N}_{0\text{d}}})$ is equal to the sum of the normal components of the displacement vector ${{\vec{u}}^\text{l}}(\vec{r},t)$, ${{\vec{u}}^\text{t}}(\vec{r},t)$ of longitudinal and transverse waves, respectively, on the plane $z=0$ ~\cite{Sen17}:
\begin{equation}
\label{eq4}
\tilde{\varsigma }({{N}_{0\text{d}}})=\left| u_{z}^\text{l}(0) \right|+\left| u_{z}^\text{t}(0) \right|.				 \end{equation}

The components $u_{z}^\text{l}(0)$, $u_{z}^\text{t}(0)$ of the displacement vector of the medium points are found from the equation solution~\cite{Pel15}:
\begin{equation}
\label{eq5}
\frac{{{\partial }^{2}}\vec{u}}{\partial {{t}^{2}}}=c_\text{t}^{2}{{\Delta }_{{\vec{r}}}}\vec{u}+(c_\text{l}^{2}-c_\text{t}^{2})\overrightarrow{\operatorname{grad}}(\operatorname{div}\vec{u}).
\end{equation}

The solution of equation (\ref{eq5}) for the Rayleigh surface wave, extending in the $x$-axis direction, is represented as:
\begin{equation}
\label{eq6}
{{u}_{x}}(x,z)=-\text{i}qA{\text{e}^{\text{i}qx-\text{i}\omega t-{{k}_\text{l}}z}}-\text{i}{{k}_\text{t}}B{\text{e}^{\text{i}qx-\text{i}\omega t-{{k}_\text{t}}z}} ,
\end{equation}
\begin{equation}
\label{eq7}
{{u}_{z}}(x,z)={{k}_\text{l}}A{\text{e}^{\text{i}qx-\text{i}\omega t-{{k}_\text{l}}z}}+qB{\text{e}^{\text{i}qx-\text{i}\omega t-{{k}_\text{t}}z}}\text{, }	
\end{equation}
where $k_\text{l,t}^{2}({{N}_{0\text{d}}})={{q}^{2}}-\frac{{{\omega }^{2}}}{c_\text{l,t}^{2}}$; \textit{A}, \textit{B} are the amplitudes of SAW.
The deformation potential created by the surface acoustic quasi-Rayleigh wave and by the adsorbed atoms is determined by the following relation:
\begin{equation}
\label{eq8}
V(x,z,t)={{\lambda }_{ij}}\frac{\partial {{u}_{i}}(x,z,t)}{\partial {{x}_{j}}}={{V}_{0}}{\text{e}^{-{{k}_\text{l}}z}}\cos \left[qx-{\omega }'(q,{{N}_{0\text{d}}})\cdot t\right],		
\end{equation}
where ${{u}_{i}}=u_{i}^\text{l}+u_{i}^\text{t}$; ${{\lambda }_{ij}}$ is a tensor of deformation potential; ${{V}_{0}}=-\frac{\left| \lambda  \right|\tilde{\varsigma }({{N}_{0\text{d}}}){{q}^{2}}(2-{{\xi }_{0}})}{{{k}_\text{l}}}\frac{c_\text{t}^{2}}{c_\text{l}^{2}};$ $\frac{1}{{{k}_\text{l}}}$ is the depth of the sound penetration into semiconductor, ${{\xi }_{0}}$ is a quantity dependent on the ratio between the longitudinal ${{c}_\text{l}}$ and transversal ${{c}_\text{t}}$ sound velocity~\cite{Pel09}, and the expression for the height of roughnesses $\tilde{\varsigma }({{N}_{0\text{d}}})$ is defined in the work~\cite{Sen17}.

\section{Electron states upon the dynamically deformed adsorbed surface}

The motion of electrons upon the dynamically deformed adsorbed surface (i.e., in the field of a quasi-Rayleigh acoustic wave) essentially depends on the value of the ratio between the wavelength ${2\piup }/{{{k}_{x}}}\;$ (${{k}_{x}}$ is the component of the electron wave vector) and the mean free path $l$ of the electrons. In semiconductors without the inversion center ${{k}_{x}}l\ll 1$, i.e., at the distances of an order of a wavelength, there is observed a multiple impact of electrons, during which the equilibrial distribution of electrons is established.
The surface electron states on the rough boundary of a monocrystal having a Zinc blende structure are found from the non-stationary Schr\"odinger equation~\cite{Sene17}:
\begin{equation}
\label{eq9}
\text{i}\hbar \frac{\partial \psi (x,z,t)}{\partial t}=-\frac{{{\hbar }^{2}}}{2{{m}^{*}}}\left( \frac{{{\partial }^{2}}\psi }{\partial {{x}^{2}}}+\frac{{{\partial }^{2}}\psi }{\partial {{z}^{2}}} \right)+{{V}_{0}}{\text{e}^{-{{k}_\text{l}}z}}\cos \left[qx-{\omega }'(q,{{N}_{0\text{d}}})\cdot t\right]\psi (x,z,t).	
\end{equation}

The surface roughnesses are formed by the dynamically deformed (acoustic quasi-Rayleigh wave) and adsorbed atoms having concentration ${{N}_{0\text{d}}}$. Interaction between the surface acoustic wave and the adsorbed atoms is attained through the deformation potential. We assume that the motion of conduction electrons is bounded by an uneven wall which is an infinitely high potential barrier.

The solution of equation (\ref{eq9}) is obtained as a sum of space-time harmonics ~\cite{Yak08}:
\begin{equation}
\label{eq10}
\psi (x,z,t)=\sum\limits_{n=-\infty }^{\infty }{{{\psi }_{n}}(z)}{\text{e}^{\text{i}{{\Phi }_{n}}(x,t)}},				 
\end{equation}
where $ {{\Phi }_{n}}(x,t)=({{k}_{x}}+nq)x-\left[{{\omega }_\text{e}}+n{\omega }'(q,{{N}_{0\text{d}}})\right]t$, $ \hbar {{\omega }_\text{e}}=E$  is the electron energy, and $\hbar {{k}_{x}}$  is its  momentum.

Substituting (\ref{eq10}) into equation (\ref{eq9}) and using the condition of orthogonality of functions, we obtain:
\begin{equation}
\label{eq11}
\left( k_{n}^{2}+\frac{{{\partial }^{2}}}{\partial {{z}^{2}}} \right){{\psi }_{n}}(z)=-2{{\beta }^{2}}q\tilde{\varsigma }({{N}_{0\text{d}}}){\text{e}^{-{{k}_\text{l}}z}}\left[{{\psi }_{n+1}}(z)+{{\psi }_{n-1}}(z)\right], \end{equation}
where $k_{n}^{2}=\frac{2{{m}^{*}}}{{{\hbar }^{2}}}\left[E+n\hbar {\omega }'(q,{{N}_{0\text{d}}})\right]-{{({{k}_{x}}+nq)}^{2}}$, ${{\beta }^{2}}=\frac{{{m}^{*}}\left| {{V}_{0}} \right|}{{{\hbar }^{2}}\tilde{\varsigma }({{N}_{0\text{d}}})q}$ is the effective electron  mass. 

Equation (\ref{eq11}) is solved using the method of successive approximations for a small parameter $q\tilde{\varsigma }({{N}_{0\text{d}}})\ll 1$. In the zero approximation, the wave function $\psi _{n}^{0}(z)$ has the form:
\begin{equation}
\label{eq12}
\psi _{n}^{0}(z)={{A}_{n}}{\text{e}^{\text{i}{{k}_{n}}z}}.				
\end{equation}

Substituting $\psi _{n}^{0}(z)$ in the right-hand part of equation~(\ref{eq11}), we find the solution of the inhomogeneous equation:
\begin{equation}
\label{eq13}
{{\tilde{\psi }}_{n}}(z)=-{{\beta }^{2}}q\tilde{\varsigma }({{N}_{0\text{d}}})\left[ \frac{{\text{e}^{\text{i}({{k}_{n-1}}+\text{i}{{k}_\text{l}})z}}}{k_{n}^{2}-{{({{k}_{n-1}}+\text{i}{{k}_\text{l}})}^{2}}}{{A}_{n-1}}-\frac{{\text{e}^{\text{i}({{k}_{n+1}}+\text{i}{{k}_\text{l}})z}}}{k_{n}^{2}-{{({{k}_{n+1}}+\text{i}{{k}_\text{l}})}^{2}}}{{A}_{n+1}} \right].
\end{equation}

The solution of Schr\"odinger equation (\ref{eq9}) can be represented as the series:
\begin{equation}
\label{eq14}
{{\psi }_{n}}(x,z,t)=\sum\limits_{n=-\infty }^{\infty }{\left[ {{A}_{n}}{\text{e}^{\text{i}{{k}_{n}}z}}-\frac{{{\beta }^{2}}q\tilde{\varsigma }({{N}_{0\text{d}}}){\text{e}^{\text{i}({{k}_{n-1}}+\text{i}{{k}_\text{l}})z}}}{k_{n}^{2}-{{({{k}_{n-1}}+\text{i}{{k}_\text{l}})}^{2}}}{{A}_{n-1}}-\frac{{{\beta }^{2}}q\tilde{\varsigma }({{N}_{0\text{d}}}){\text{e}^{\text{i}({{k}_{n+1}}+\text{i}{{k}_\text{l}})z}}}{k_{n}^{2}-{{({{k}_{n+1}}+\text{i}{{k}_\text{l}})}^{2}}}{{A}_{n+1}} \right]}{\text{e}^{\text{i}{{\Phi }_{n}}(x,t)}}.
\end{equation}

Finding the dispersion law of surface electron states $E=E({{k}_{x}})$ we use the boundary conditions for the wave function at infinity and at the interface boundary. Taking into account the conditions for a smooth roughness of the surface $q\tilde{\varsigma }({{N}_{0\text{d}}})\ll 1$ $\big(\frac{\partial {{z}_{0}}}{\partial x}\ll 1 \big)$, the boundary condition on the plane $z=0$ ~\cite{Eme02} takes the form~\cite{Pel17}:
\begin{equation}
\label{eq15}
\frac{\partial \psi }{\partial z}+\frac{{{\partial }^{2}}\psi }{\partial {{z}^{2}}}{{z}_{0}}-{{\left. \frac{\partial \psi }{\partial x}\frac{\partial {{z}_{0}}}{\partial x} \right|}_{z=0}}=0.			
\end{equation}

At a distribution of a surface acoustic wave the  electrons on the surface scatter on its deformation potential $V(x,z,t)$~\cite{Sene17} within the volume, as well as on the roughnesses which are formed by both the surface acoustic wave and the heterogeneous distribution of adsorbed atoms. As a result, the harmonics are actuated with wave numbers ${{k}_{n+1}}$  and ${{k}_{n-1}}$.
Within the boundary condition (\ref{eq15}) we substitute the wave function in the form~(\ref{eq14})~\cite{Sene17}. In the following calculations, the energy of the Rayleigh wave quantum is neglected, i.e.:
\begin{center}
$\hbar {\omega }'(q,{{N}_{0\text{d}}})\ll \frac{{{\hbar }^{2}}{{q}^{2}}}{2{{m}^{*}}}$
\end{center}
or
       \begin{equation}
\label{eq16}
       {{\xi }_{0}}{{c}_\text{t}}\ll\frac{\hbar q}{2{{m}^{*}}}.			
\end{equation}

If the surface roughness is formed by a quasi-Rayleigh wave, then the electrons dissipate not only on the roughness created by the inhomogeneous distribution of adsorbed atoms ${{N}_\text{d}}(x)$ but also on the deformation potential of the quasi-Rayleigh wave. This additional scattering prevents the formation of surface electron states.

Now, we consider the existence of surface electron states for a zero harmonic. In this case, the interaction with the harmonic $n\geqslant 2$ is neglected, because the connection with the zero harmonic is proportional to ${{\left[q\tilde{\varsigma }({{N}_{0\text{d}}})\right]}^{n}}\ll 1$. The spectrum of electron states on a surface with roughnesses, formed by the surface acoustic wave and by adsorbed atoms, has the form~\cite{Sene17}:
\begin{eqnarray}
\label{eq17}
   &&{{k}_{0}}=-\frac{{{{\tilde{\varsigma }}}^{2}}({{N}_{0\text{d}}})}{4}\left[ \frac{{{\left(k_{0}^{2}+{{k}_{x}}q\right)}^{2}}}{{{k}_{-1}}}+\frac{{{\left(k_{0}^{2}-{{k}_{x}}q\right)}^{2}}}{{{k}_{1}}} \right]-\frac{\text{i}{{\beta }^{2}}q{{{\tilde{\varsigma }}}^{2}}({{N}_{0\text{d}}})}{2} \nonumber \\
   &&\times \left\{ \frac{k_{0}^{2}+{{k}_{x}}q}{{{k}_{-1}}}\left[ \frac{{{k}_{0}}+\text{i}{{k}_\text{l}}}{k_{-1}^{2}-{{({{k}_{0}}+\text{i}{{k}_\text{l}})}^{2}}}+\frac{{{k}_{-1}}+\text{i}{{k}_\text{l}}}{k_{0}^{2}-{{({{k}_{-1}}+\text{i}{{k}_\text{l}})}^{2}}} \right]+ \right.\frac{k_{0}^{2}-{{k}_{x}}q}{{{k}_{1}}} \nonumber \\
   &&\times \left. \left[ \frac{{{k}_{0}}+\text{i}{{k}_\text{l}}}{k_{1}^{2}-{{({{k}_{0}}+\text{i}{{k}_\text{l}})}^{2}}}+\frac{{{k}_{1}}+\text{i}{{k}_\text{l}}}{k_{0}^{2}-{{({{k}_{1}}+\text{i}{{k}_\text{l}})}^{2}}} \right]-{{k}_{0}}\left( \frac{{{\Gamma }_{1}}}{{{k}_{1}}}+\frac{{{\Gamma }_{-1}}}{{{k}_{-1}}} \right)+{{\Gamma }_{0}} \right\},
\end{eqnarray}
where		${{\Gamma }_{0}}=\frac{q({{k}_{x}}-q)-{{({{k}_{0}}+\text{i}{{k}_\text{l}})}^{2}}}{k_{-1}^{2}-{{({{k}_{0}}+\text{i}{{k}_\text{l}})}^{2}}}-\frac{q({{k}_{x}}+q)+{{({{k}_{0}}+\text{i}{{k}_\text{l}})}^{2}}}{k_{1}^{2}-{{({{k}_{0}}+\text{i}{{k}_\text{l}})}^{2}}},$
${{\Gamma }_{\pm 1}}=\frac{({{k}_{\pm 1}}+\text{i}{{k}_\text{l}})\mp {{k}_{x}}q}{k_{0}^{2}-{{({{k}_{\pm 1}}+\text{i}{{k}_\text{l}})}^{2}}}.$

 We find the solution of equation (\ref{eq17}) using the method of successive approximations ${{k}_{0}}\left(\tilde{\varsigma }({{N}_{0\text{d}}})\right)={{k}_{0}}\left(\tilde{\varsigma }(0)\right)+\delta {{k}_{0}}\left(\tilde{\varsigma }({{N}_{0\text{d}}})\right)$ for a small parameter $q\tilde{\varsigma }({{N}_{0\text{d}}})\ll 1$ for amplitudes with $n=1;0;-1$, where $\tilde{\varsigma }({{N}_{0\text{d}}})=\tilde{\varsigma }(0)+\delta \tilde{\varsigma }({{N}_{0\text{d}}})$, $\tilde{\varsigma }(0)$ is the height of the roughnesses of the surface formed only by a surface acoustic wave.
For the case $\tilde{\varsigma }({{N}_{0\text{d}}})=\tilde{\varsigma }_{0}^{{}}+\delta \tilde{\varsigma }({{N}_{0\text{d}}})=0$ (the absence of the adsorbed atoms and a surface acoustic wave, i.e., the case of a smooth surface), the solution of equation (\ref{eq17}) is ${{k}_{0}}\equiv k_{0}^{(0)}=0$. Then, the dispersion law of electrons moving along a smooth surface has the form ${{E}_{0}}=\frac{{{\hbar }^{2}}k_{x}^{2}}{2{{m}^{*}}}$. In this case, the region where the electrons localize ${{L}_\text{e}}({{N}_{0\text{d}}})={1}/{\left| \delta {{k}_{0}}({{N}_{0\text{d}}}) \right|}$ occupies a half of $x\in (0;\infty )$ space, since $\left| \delta {{k}_{0}}({{N}_{0\text{d}}}) \right|\to 0$.
At $\tilde{\varsigma }\ne 0$, the zero harmonic ($n=0$)  $k_{0}^{2}=\frac{2{{m}^{*}}}{{{\hbar }^{2}}}E-k_{x}^{2}.$ Then,
\begin{equation}
\label{eq18}
E=\frac{{{\hbar }^{2}}}{2{{m}^{*}}}\big(k_{0}^{2}+k_{x}^{2}\big)=\frac{{{\hbar }^{2}}k_{x}^{2}}{2{{m}^{*}}}\Bigg( 1+\frac{k_{0}^{2}}{k_{x}^{2}} \Bigg).
\end{equation}
Let ${{k}_{0}}=k_{0}^{(0)}+\delta {{k}_{0}}$. Then,
\begin{equation}
\label{eq19}
E=\frac{{{\hbar }^{2}}k_{x}^{2}}{2{{m}^{*}}}\left[ 1+\frac{{{(\delta {{k}_{0}})}^{2}}}{k_{x}^{2}} \right],\,\,\, k_{0}^{(0)}=0; \nonumber			
\end{equation}
\begin{eqnarray}
&&\delta {{k}_{0}}=-\frac{{{{\tilde{\varsigma }}}^{2}}({{N}_{0\text{d}}})k_{x}^{2}{{q}^{2}}}{4}\left( \frac{1}{{{k}_{1}}}+\frac{1}{{{k}_{-1}}} \right)-\frac{\text{i}{{\beta }^{2}}q{{{\tilde{\varsigma }}}^{2}}({{N}_{0\text{d}}})}{2}  
\Bigg\{ \big({{q}^{2}}-k_{l}^{2}\big)\left( \frac{1}{k_{1}^{2}+{{k}_\text{l}}^{2}}+\frac{1}{k_{-1}^{2}+{{k}_\text{l}}^{2}} \right)\nonumber \\
&&+2\text{i}{{k}_\text{l}}{{k}_{x}}q \left[ \frac{1}{{{k}_{1}}\big(k_{1}^{2}+{{k}_\text{l}}^{2}\big)}-\frac{1}{{{k}_{-1}}\big(k_{-1}^{2}+{{k}_\text{l}}^{2}\big)} \right] \Bigg\}.
\end{eqnarray}

Now, we find the quantity $\delta {{k}_{0}}$ in the boundary cases: long-wave (${{k}_{x}}\ll {q}/{2}\;$), resonance (${{k}_{x}}\sim {q}/{2}\;$) and short-wave (${{k}_{x}}\gg {q}/{2}\;$) cases.

In the long-wave approximation, the value $\delta {{k}_{0}}$ has the form:
\begin{equation}
\label{eq20}
\delta {{k}_{0}}=\big(k_{x}^{2}-2{{\beta }^{2}}\big)\frac{{{q}^{2}}}{4}\left( \frac{1}{{{k}_{1}}}+\frac{1}{{{k}_{-1}}} \right)\left[\tilde{\varsigma }_{0}^{2}+2{{\tilde{\varsigma }}_{0}}\delta \tilde{\varsigma }({{N}_{0\text{d}}})\right],		
\end{equation}
where  ${{k}_{1}}=\text{i}\sqrt{q(q+2{{k}_{x}})}$,  ${{k}_{-1}}=\text{i}\sqrt{q(q-2{{k}_{x}})}$.

In the resonance case, the dispersion equation has the form:

$$\delta {{k}_{0}}\delta {{k}_{-1}}=-\frac{\tilde{\zeta }_{0}^{2}+2{{{\tilde{\zeta }}}_{0}}\delta \tilde{\zeta }({{N}_{0\text{d}}})}{16}{{q}^{4}}\left( 1-\frac{8{{\beta }^{2}}}{q{{k}_\text{l}}} \right),$$
where $\delta {{k}_{0}}=\sqrt{\frac{2{{m}^{*}}}{{{\hbar }^{2}}}\delta E-q\delta {{k}_{x}}}$, $\delta {{k}_{-1}}=\sqrt{\frac{2{{m}^{*}}}{{{\hbar }^{2}}}\delta E+q\delta {{k}_{x}}}$.

At the point ${{k}_{x}}=\frac{q}{2}\big( {{k}_{x}}=\frac{\piup }{{{L}_{x}}} \big) $, the energy changes jump-like, i.e., there is a band gap whose value is equal to
\begin{equation}
\label{eq21}
2\delta E=-\frac{{{\hbar }^{2}}{{q}^{4}}}{16{{m}^{*}}}\left( 1-\frac{8{{\beta }^{2}}}{q{{k}_\text{l}}} \right)\left[ \tilde{\zeta }_{0}^{2}+2{{{\tilde{\zeta }}}_{0}}\delta \tilde{\zeta }({{N}_{0\text{d}}}) \right].		
\end{equation}

As can be seen from formula (\ref{eq21}), in the case of the absence of adsorbed atoms  $\left[ \delta \tilde{\zeta }({{N}_{0\text{d}}})=0 \right]$, the width of the band gap in the subsurface layer of the GaAs(100) semiconductor coincides with the results of work~\cite{Yak08}.

In the short-wave approximation, the values  $ \delta {{k}_{0}}$ and $E$ are complex, i.e., the electron states are quasi-stationary.
\begin{equation}
E=\operatorname{Re}E+\text{i}\operatorname{Im}E, \nonumber 
\end{equation}
\begin{equation}
\operatorname{Re}E=\frac{{{\hbar }^{2}}k_{x}^{2}}{2{{m}^{*}}}\left[ 1+\frac{{{(\operatorname{Re}\delta {{k}_{0}})}^{2}}-{{(\operatorname{Im}\delta {{k}_{0}})}^{2}}}{k_{x}^{2}} \right], \nonumber
\end{equation}
\begin{equation}
\operatorname{Im}E=\frac{{{\hbar }^{2}}}{2{{m}^{*}}}\operatorname{Re}\delta {{k}_{0}}\operatorname{Im}\delta {{k}_{0}}<0, \nonumber
\end{equation}
where		
\begin{align}
 \operatorname{Re}\delta {{k}_{0}}&=-\frac{{{q}^{2}}}{{{k}_{-1}}}\left[ \tilde{\zeta }_{0}^{2}+2{{{\tilde{\zeta }}}_{0}}\delta \tilde{\zeta }({{N}_{0\text{d}}}) \right]\big( k_{x}^{2}+2{{\beta }^{2}} \big), \nonumber
\\
\operatorname{Im}\delta {{k}_{0}}&=-\frac{{{q}^{2}}}{4\left| {{k}_{1}} \right|}\left[ \tilde{\zeta }_{0}^{2}+2{{{\tilde{\zeta }}}_{0}}\delta \tilde{\zeta }({{N}_{0\text{d}}}) \right]\big( k_{x}^{2}-2{{\beta }^{2}} \big), \nonumber
\\
{{k}_{1}}&=\text{i}\sqrt{q(q+2{{k}_{x}})}\,,\,\,\,\, {{k}_{-1}}=\sqrt{q(2{{k}_{x}}-q)}. \nonumber
\end{align}

At the same time, the relaxation time is equal to $\tau =\hbar /\left| \operatorname{Im}E \right| $.

\section{Influence of the concentration of adsorbed atoms on the band gap width and the electron density distribution on the surface of a mo\-no\-crys\-tal having a Zinc blende structure}

\begin{figure}[!b]
   \begin{center}
   \includegraphics[width=300pt]{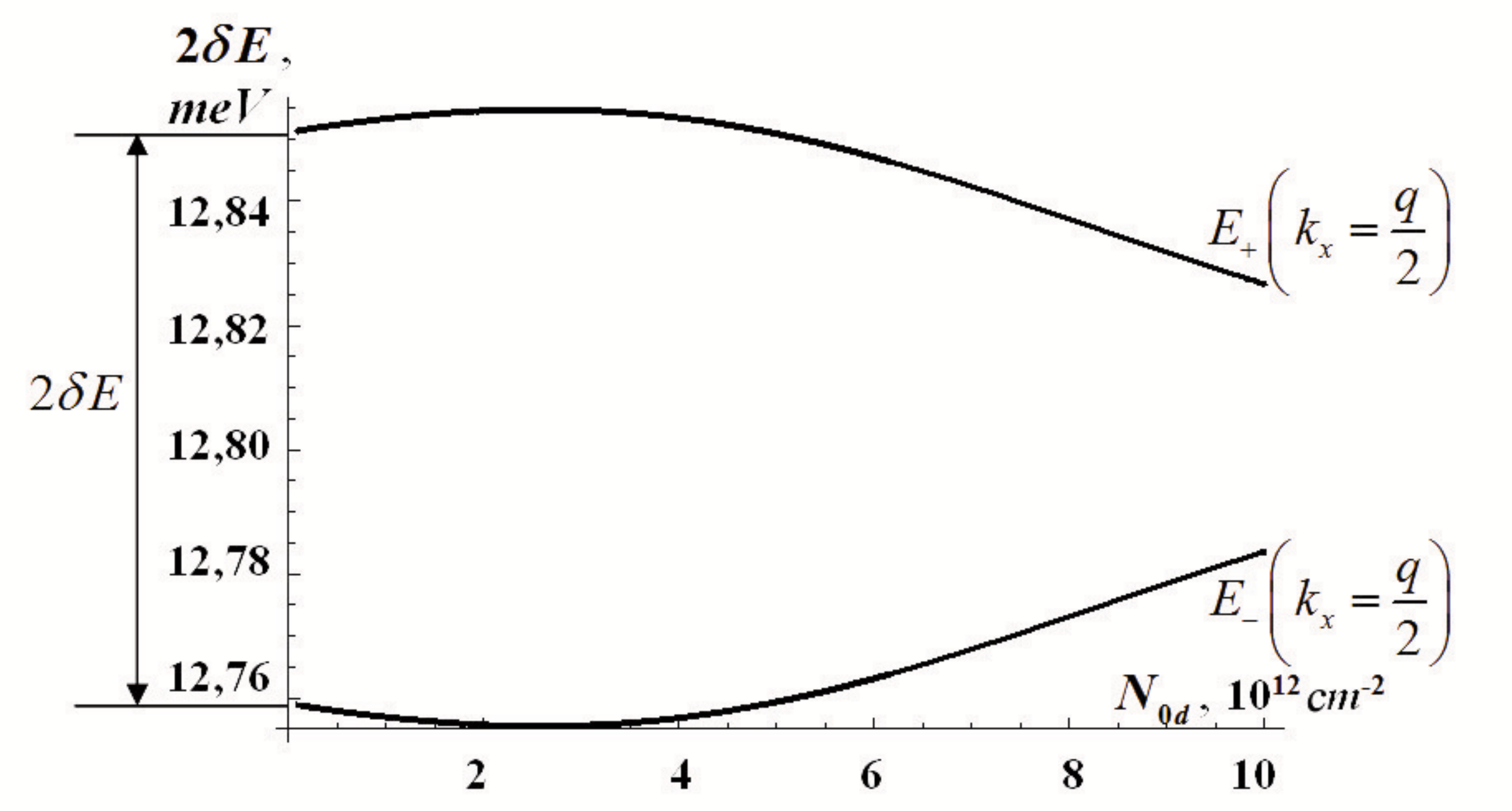}
   \caption{The band gap width on the surface of GaAs(100) semiconductor at the point ${{k}_{x}}={q}/{2}\;$ of the Brillouin zone, depending on the concentration of adsorbed atoms.}
   \label{fig1}
   \end{center}
\end{figure}

Figure~\ref{fig1} shows a plot of the dependence of the band gap width on the concentration of adsorbed atoms upon the plane of a monocrystal having a Zinc blende structure at the point of the Brillouin zone  ${{k}_{x}}={q}/{2}\;$ or ${{k}_{x}}=\piup /{{L}_{x}}$. The upper and lower branches are determined according to the relations~\cite{Sene17}:
\begin{eqnarray}
\label{eq22}
{{E}_{+}}\left( {{k}_{x}}=\frac{q}{2} \right)=-\frac{{{\hbar }^{2}}{{q}^{2}}}{8{{m}^{*}}}\left\{ 1+\frac{{{q}^{2}}}{4}\left( 1-\frac{8{{\beta }^{2}}}{q{{k}_\text{l}}} \right)\left[ \tilde{\zeta }_{0}^{2}+2{{{\tilde{\zeta }}}_{0}}\delta \tilde{\zeta }({{N}_{0\text{d}}}) \right] \right\},
\end{eqnarray}
\begin{eqnarray}
\label{eq23}
{{E}_{-}}\left( {{k}_{x}}=\frac{q}{2} \right)=-\frac{{{\hbar }^{2}}{{q}^{2}}}{8{{m}^{*}}}\left\{ 1-\frac{{{q}^{2}}}{4}\left( 1-\frac{8{{\beta }^{2}}}{q{{k}_\text{l}}} \right)\left[ \tilde{\zeta }_{0}^{2}+2{{{\tilde{\zeta }}}_{0}}\delta \tilde{\zeta }({{N}_{0\text{d}}}) \right] \right\}.		
\end{eqnarray}

The calculation was made for GaAs(100) semiconductor with the following parameter values ~\cite{Pel15, Eme02}:
${{l}_\text{d}}=2.9$ nm; $a=0.565$ nm; ${{c}_\text{l}}=4400$ m/s; ${{c}_\text{t}}=2475$ m/s; $\rho =5320$ kg/m$^{3}$; ${{\theta }_\text{d}}=10$ eV; ${{N}_{0\text{d}}}=3\cdot {{10}^{13}}$~cm$^{-2}$; ${{D}_\text{d}}=5\cdot {{10}^{-2}}$ cm$^{2}$/s; $\frac{\partial F}{\partial {{N}_\text{d}}}=0.1$ eV; ${{m}^{*}}=6.1\cdot {{10}^{-32}}$ kg; $\lambda =0.02$ eV.

Figure~\ref{fig1} shows that the functional dependence of the band gap width on the concentration of adsorbed atoms is nonmonotonous.
There is observed an increase of a functional dependence $\delta E=\delta E({{N}_{0\text{d}}})$ on the concentration interval of adatoms $0<{{N}_{0\text{d}}}\leqslant 2.1\cdot {{10}^{12}}$ ${\text{cm}^{-2}}$.
At this concentration interval, the semiconductor band gap width increases by 8\%.
With a further increase of the concentration of adsorbed atoms, the band gap width decreases. Moreover, at the interval $2.1\cdot {{10}^{12}}<{{N}_{0\text{d}}}\leqslant {{10}^{13}}$ ${\text{cm}^{-2}}$, the band gap width decreases by 56\%.
Such nonmonotonous dependence of $\delta E=\delta E({{N}_{0\text{d}}})$ is explained by the nonmonotonous dependence of the roughness height $\tilde{\varsigma }=\tilde{\varsigma }({{N}_{0\text{d}}})$ on the concentration of adsorbed atoms~\cite{Sene17}.

The length of the spatial localization of an electron wave function ${{L}_\text{e}}={{L}_\text{e}}({{N}_{0\text{d}}})$ in the long-wave approximation is obtained by the relation:
\begin{equation}
\label{eq24}
{{L}_{}}({{N}_{0\text{d}}})=\frac{4}{{{q}^{2}}\left( \frac{1}{{{k}_{1}}}+\frac{1}{{{k}_{-1}}} \right)\left[\tilde{\varsigma }_{0}^{2}+2{{{\tilde{\varsigma }}}_{0}}\delta \tilde{\varsigma }({{N}_{0\text{d}}})\right]\big(k_{x}^{2}-2{{\beta }^{2}}\big)}.      		
\end{equation}

We can see from formula (\ref{eq24}), that the reduction of the roughness period ${{L}_{x}}$  $\big(q=\frac{2\piup }{{{L}_{x}}}\big)$ along the $x$-axis or an increase of the height $\tilde{\varsigma }({{N}_{0\text{d}}})=\tilde{\varsigma }_{0}^{{}}+\delta \tilde{\varsigma }({{N}_{0\text{d}}})$ of the adsorbed surface roughness leads to a stronger localization of the electron wave function
${{\psi }_{0}}={{A}_{0}}{\text{e}^{\text{i}{{k}_{x}}x-{z}/{L}({{N}_{0\text{d}}})}}.$
A decrease of the length of an electron de Broglie wave (an increase of ${{k}_{x}}$) reduces to the same effect.

\begin{figure}[!b]
   \begin{center}
   \includegraphics[width=300pt]{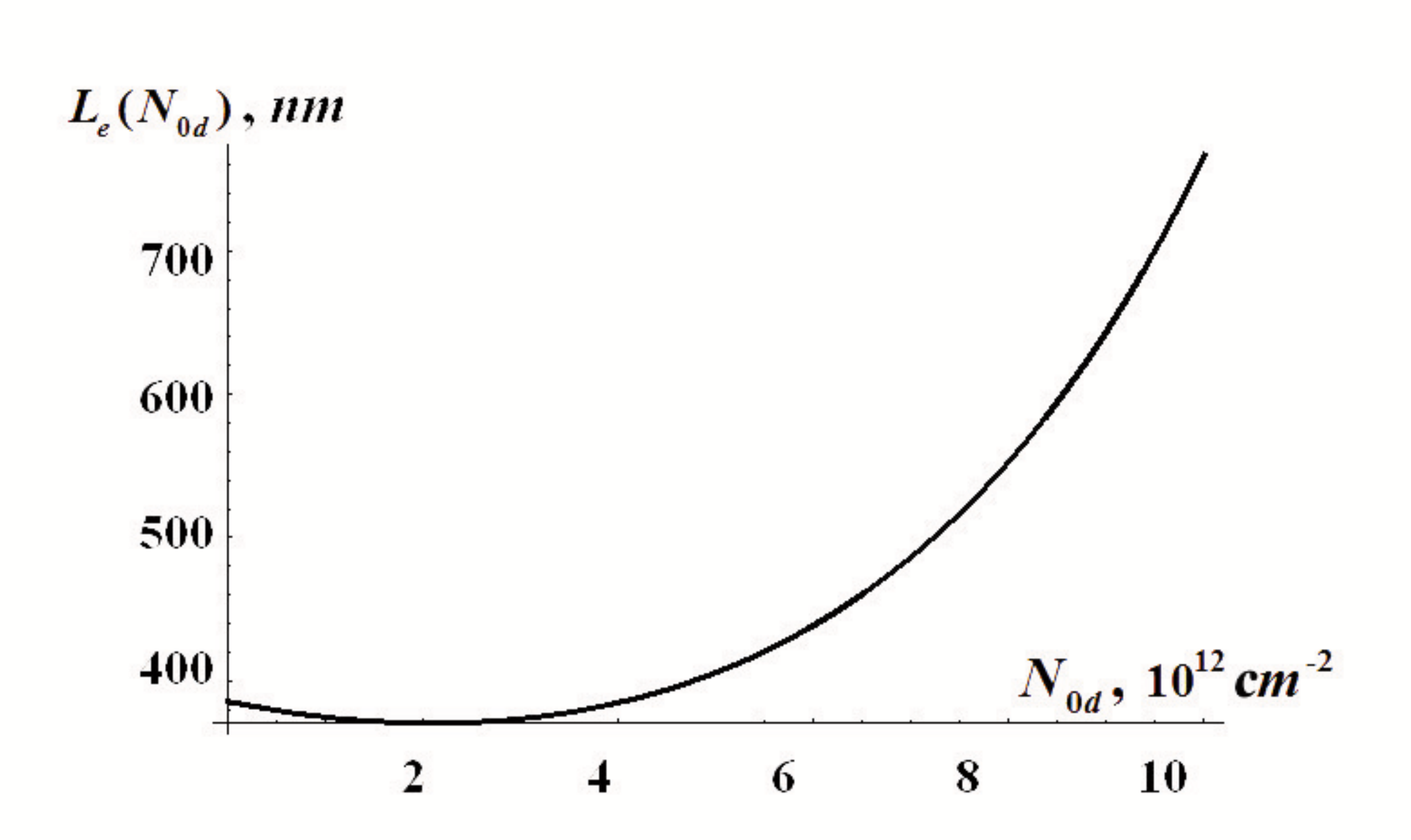}
   \caption{The length of the spatial localization of an electron wave function in the subsurface of GaAs(100) semiconductor, depending on the concentration of adsorbed atoms at $ k_x=q/3$;  $q=0.03$~\AA$^{-1}$.}
   \label{fig2}
   \end{center}
\end{figure}

Figure~\ref{fig2} shows the dependence of the length of a spatial localization of the electron wave function ${{L}_\text{e}}={{L}_\text{e}}({{N}_{0\text{d}}})$ on the concentration of adsorbed atoms in the case of long-wave approximation.

Analyzing the graphic dependence in figure~\ref{fig2}, we see that curve ${{L}_\text{e}}={{L}_\text{e}}({{N}_{0\text{d}}})$ is nonmonotonous. At small concentrations of adsorbed atoms ($0<{{N}_{0\text{d}}}\leqslant 2.1 \cdot {{10}^{12}}$ ${\text{cm}^{-2}}$) with an increasing concentration, the length of the spatial localization of the wave function of the electron decreases. With a further increase of the concentration of adsorbed atoms, the function ${{L}_\text{e}}={{L}_\text{e}}({{N}_{0\text{d}}})$ increases monotonously~\cite{Pel16}. The roughness of the two media division is caused by the appearance of surface electron states.

The wave function of the surface electronic states is determined by the following expression~\cite{Kha11}
${{\psi }_{k}}={{A}_{0}}{\text{e}^{-\left| \delta {{k}_{z}} \right|z+\text{i}\left| {{k}_{x}}x-{{\omega }_{\vec{k}}}t \right|}}$.
From the normalization condition
$\iiint{{{\psi }_{k}} {\psi^{*}_{k}}}\rd x \rd y \rd z=1$,
we determine the amplitude where ${{A}_{0}}=\sqrt{{2\left| \delta {{k}_{z}} \right|}/{S}\;}$, where $S={{L}_{x}}{{L}_{y}}$; ${{L}_{x}},{{L}_{y}}$ are the model sizes in $x$ and $y$ directions, respectively.

The dependence of the wave function of surface electron states on the \textit{z} coordinate leads to the appearance of an inhomogeneous electron plasma in the $z>z_0$ region. The basic parameters of this plasma can be determined due to the characteristics of the surface roughness. We determine the change of the concentration of electrons $\delta {{n}_{0}}(z)$ as:
$$\delta {{n}_{0}}(z)=\sum\limits_{{{k}_{x}}}{{{\psi }_{k}}\psi _{k}^{*}{{n}_{{{k}_{x}}}}=\frac{2}{S}}\sum\limits_{{{k}_{x}}}{{{n}_{{{k}_{z}}}}\left| \delta {{k}_{0}} \right|}{\text{e}^{-2\left| \delta {{k}_{z}} \right|z}},$$
where ${{n}_{{{k}_{x}}}}$ is the number of electrons with the wave vector ${{k}_{x}}$; the sum is carried out at all values of ${{k}_{x}}$. In this case, the minimum ${{k}_{x}}$ value is determined by the model sizes in the $x$-direction, that is, the value~${{L}_{x}}$, and the maximum is determined by the Fermi  momentum $\hbar {{k}_\text{F}}$. The total number of particles in the $z>0$ region is equal to  $\sum\nolimits_{{{k}_{x}}}{{{n}_{{{k}_{x}}}}}$.

\begin{figure}[!t]
   \begin{center}
   \includegraphics[width=300pt]{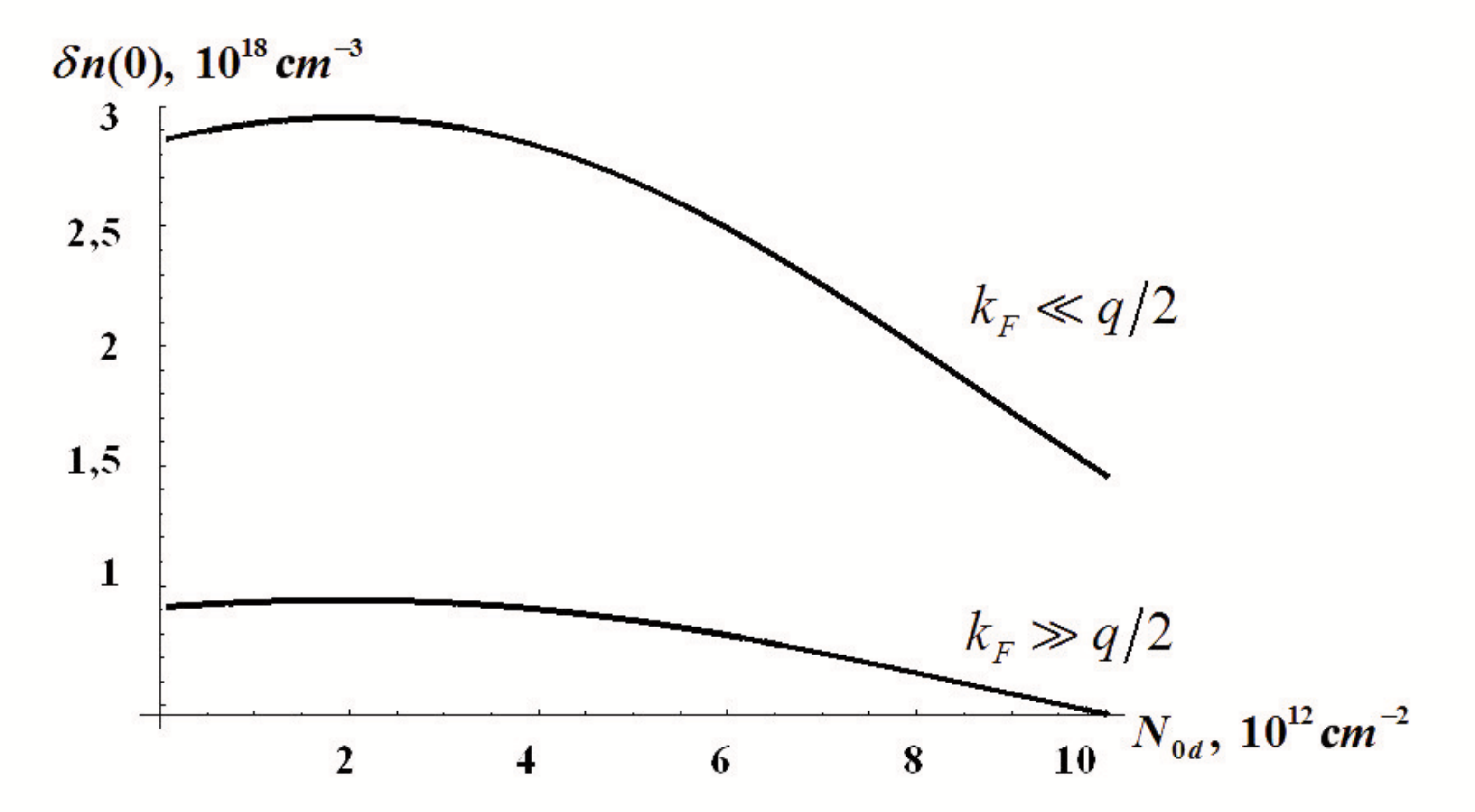}
   \caption{The change of the  concentration of electrons upon the $z=0$ surface of the GaAs(100) semiconductor, depending on the concentration of adsorbed atoms.}
   \label{fig3}
   \end{center}
\end{figure}

The surface density
${{n}_{S}}=\int\nolimits_{0}^{\infty }{\delta {{n}_{0}}(z)\rd z}=\sum\nolimits_{{{k}_{x}}}{{{n}_{{{k}_{x}}}}}/S.$
For a degenerate electron gas at ${{n}_{{{k}_{x}}}}=0$ and ${{n}_{{{k}_{x}}}}=1$, the values $\delta {{n}_{0}}(0)=2{{n}_{s}}\left| \delta {{k}_{0}} \right|$ are as follows:

-	if ${{k}_\text{F}}<{q}/{2}\;$ then  $\delta {{n}_{0}}(0)=2\left| \delta {{k}_{0}} \right|\frac{{{k}_\text{F}}^{2}}{4\piup }$;

-	if ${{k}_\text{F}}>{q}/{2}\;$ then $\delta {{n}_{0}}(0)=2\left| \delta {{k}_{0}} \right|\frac{{{k}_\text{F}}q}{2{{\piup }^{2}}}$.

A graphic representation of the dependence of the change of the concentration of electrons $\delta n(0)$ on the  concentration of adsorbed atoms upon the $z=0$ surface is given in figure~\ref{fig3}. A nonmonotonous character of the curves is determined by a nonmonotonous functional dependence of the height of the roughnesses of semiconductor surfaces on the concentration of adsorbed atoms~\cite{Sen17}.

The maximum change of the concentration of electrons upon $z=0$ surface of GaAs(100) semiconductor is observed at ${{N}_{0\text{d}}}=1.9\cdot {{10}^{12}}$ $\text{cm}^{-2}$. In particular, if ${{k}_\text{F}}>{q}/{2}$  (${{k}_\text{F}}=0.06$~\AA$^{-1}$; $q/2=0.015$~\AA$^{-1}$), then $\delta n(0)=9.4\cdot {{10}^{17}}$ $\text{cm}^{-3}$; if  ${{k}_\text{F}}<{q}/{2}$ ~(${{k}_\text{F}}=0.006$~\AA$^{-1}$) then $\delta n(0)=2.95\cdot {{10}^{18}}~\text{cm}^{-3}$. As the concentrations of adsorbed atoms increases (${{N}_{0\text{d}}}>1.9\cdot {{10}^{12}}~\text{cm}^{-2}$), the change of the electron density distribution upon the surface of GaAs(100) semiconductor monotonously decreases.

\section{Conclusions}

\begin{enumerate}
\item The dispersion relations for the spectra of surface electron states upon a dynamically deformed adsorbed surface of GaAs (CdTe) semiconductor were derived in the long-wave (${{k}_{x}}\ll {q}/{2}$), resonant (${{k}_{x}}\sim{q}/{2}$) and short-wave (${{k}_{x}}\gg {q}/{2}$) approximations.
\item	It was established that the energy band gap width upon a dynamically deformed adsorbed surface of a semiconductor at the point ${{k}_{x}}={q}/{2}\;$ of the Brillouin zone, depending on the concentration of adsorbed atoms, is of a nonmonotonous character.
\item	It was shown that the dependence of the change of the concentration of electrons on the concentration of adsorbed atoms upon the surface is of a nonmonotonous character, determined by the height of the inequality of the solid surface. In particular, the maximum change of the electron density upon the surface of a GaAs(100) semiconductor  is achieved at the concentration of the adsorbed atoms ${{N}_{0\text{d}}}=1.9\cdot {{10}^{12}}$ $\text{cm}^{-2}$.
\end{enumerate}

%
%

\ukrainianpart

\title{Теорія електронних станів на динамічно деформованій адсорбованій поверхні твердого тіла}
\author[Р.М.~Пелещак, М.Я.~Сенета \ldots]{Р.М.~Пелещак, М.Я.~Сенета}
\address{Дрогобицький державний педагогічний університет імені Івана Франка, \\ вул. Івана Франка, 24, 82100 Дрогобич, Україна
}
%
%
%

\makeukrtitle

\begin{abstract}
\tolerance=3000%
Отримано дисперсійні співвідношення для спектру поверхневих електронних станів на динамічно деформованій адсорбованій поверхні монокристалу зі структурою цинкової обманки. Встановлено, що залежності ширини забороненої зони та концентрації електронів на поверхні твердого тіла від концентрації адсорбованих атомів ${{N}_{0\text{d}}}$ мають немонотонний характер
\keywords електронні стани, акустична квазірелеєвська хвиля, адсорбовані атоми

\end{abstract}

\end{document}